\documentclass[aps,twocolumn,groupedaddress]{revtex4}
\usepackage{epsf,latexsym}
\usepackage{times}

\newcommand{\ket}[1]{| #1 \rangle}

\newcommand{\ra}{{\rightarrow}}
\newcommand{\up}{{\uparrow}}
\newcommand{\down}{{\downarrow}}

\newcommand{\be}{\begin{equation}}
\newcommand{\ee}{\end{equation}}

\def\CC{{\rm\kern.24em \vrule width.04em height1.46ex depth-.07ex
    \kern-.30em C}}
\def\P{{\rm I\kern-.25em P}}

\def\bbbc{{\mathchoice {\setbox0=\hbox{$\displaystyle\rm C$}\hbox{\hbox
to0pt{\kern0.4\wd0\vrule height0.9\ht0\hss}\box0}}
{\setbox0=\hbox{$\textstyle\rm C$}\hbox{\hbox
to0pt{\kern0.4\wd0\vrule height0.9\ht0\hss}\box0}}
{\setbox0=\hbox{$\scriptstyle\rm C$}\hbox{\hbox
to0pt{\kern0.4\wd0\vrule height0.9\ht0\hss}\box0}}
{\setbox0=\hbox{$\scriptscriptstyle\rm C$}\hbox{\hbox
to0pt{\kern0.4\wd0\vrule height0.9\ht0\hss}\box0}}}}

\def\bbbz{{\mathchoice {\hbox{$\sf\textstyle Z\kern-0.4em Z$}}
{\hbox{$\sf\textstyle Z\kern-0.4em Z$}}
{\hbox{$\sf\scriptstyle Z\kern-0.3em Z$}}
{\hbox{$\sf\scriptscriptstyle Z\kern-0.2em Z$}}}}

\newcommand{\putfig}[2]{$$\leavevmode\hbox{\epsfxsize=#2 cm
   \epsffile{#1.eps}}$$}

\begin{document}
\title{Mesoscopic Stern-Gerlach device to polarize spin currents}
\author{Radu Ionicioiu}
\author{Irene D'Amico}
\affiliation{Institute for Scientific Interchange (ISI), Villa Gualino, Viale Settimio Severo 65, I-10133 Torino, Italy}

\begin{abstract}
Spin preparation and spin detection are fundamental problems in spintronics and in 
several solid state proposals for quantum information processing. Here we propose the 
mesoscopic equivalent of an optical polarizing beam splitter (PBS). This interferometric 
device uses non-dispersive phases (Aharonov-Bohm and Rashba) in order to separate spin up 
and spin down carriers into distinct outputs and thus it is analogous to a Stern-Gerlach apparatus. 
It can be used both as a spin preparation device and as a spin measuring device by converting 
spin into charge (orbital) degrees of freedom. An important feature of the proposed spin
 polarizer is that no ferromagnetic contacts are used.
\end{abstract}

\maketitle

One of the most important problems in the newly emerged field of spintronics 
\cite{spintronics} is to design and build a controlled source of spin polarized electrons, in particular when semiconductor-based micro-circuits are considered.
 One possibility is to inject electrons from a ferromagnetic contact, like in the pioneering proposal of Datta and Das \cite{datta_das}. However, this approach presents some intrinsic obstacles, related mainly to the
 conductivity mismatch between metals and semiconductors \cite{Schmidt:2000, Schmidt:2002, Flatte:20022002, Rashba}.
Spin-injection rates of a few percents were in fact
 reported in experiments based on ferromagnet/semiconductor 
junctions \cite{spin_injection, gardelis}, though a higher spin polarized current has been injected in GaAs
 using a ferromagnetic scanning tunneling microscope tip \cite{LaBella-Byers}. 
A partial solution to the injection problem (at least at low temperatures)
is to use magnetic-semiconductor/semiconductor interfaces \cite{spin_injection2, ohno, jonker}, for which
polarization rates as high as 90\% have been achieved \cite{spin_injection2}. Finally, an additional concern related mainly to spin-based 
quantum computation devices, is the requirement of a high degree of control on the single-spin dynamics and coherence as well as the possibility of single-spin detection. 
In order to overcome some of the aforementioned problems, several 
mechanisms for spin polarizing devices and filters have been recently 
proposed \cite{t_spin, spin_filter, spin_switch, QDspin_filter, nitta, t_spin1}.

In this article we present a mesoscopic device equivalent to the optical polarizing beam splitter (PBS). Such device can be used both to {\it inject} spin polarized electrons, with an efficiency as high as 100\%, and to {\it detect}
 single electron spins. We will show that our scheme is robust against a large class of perturbations and we will discuss the relevant parameter regimes.

A PBS is a four terminal device with two inputs (called {\em modes} and 
labeled 0 and 1, respectively) and two outputs (0' and 1'). For each input, it transmits one 
polarization into the same mode and reflects the other polarization into the opposite mode. 
Thus, an incoming spin up (down) in mode 0 is transmitted (reflected) to the 
output mode 0' (1') (and similarly for spins incoming in the 1-mode). Let $\sigma= \up, \down$
 be the spin degrees of freedom and $k=0, 1$ the orbital degrees of freedom (the modes). Then 
a PBS realizes the following unitary transformation: $\ket{\up;\, k} \ra\ \ket{\up;\, k}$, 
$\ket{\down; k} \ra\ \ket{\down; 1-k}$. In the terminology of quantum gates, a PBS is a 
{\sf CNOT} gate in which the spin acts as a 'control' for the orbital degrees of freedom 
(the 'target'). Using only one active input (say 0), an unpolarized beam of incoming spins 
is separated into two completely polarized outputs, and the device is thus equivalent to a 
Stern-Gerlach apparatus \cite{oneinput}.

\begin{figure}
\putfig{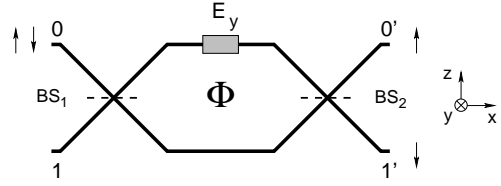}{6.5}
\caption{A sketch of the proposed spin polarizing beam splitter. Unpolarized spins are injected in mode 0; the 0' (1') output mode contains only spin up (down) polarized electrons. BS$_{1,2}$ are two beam splitters; a magnetic flux $\Phi$ is applied through the interferometer generating an Aharonov-Bohm phase. An electric field $E_y$ is applied locally on the upper arm (the 1-mode) generating a Rashba phase (the gray box).}
\label{pbs}
\end{figure}

Our four-terminal device is a Mach-Zehnder interferometer (MZI) with a local spin-orbit 
(Rashba) interaction on the upper arm (the 1-mode) and a global magnetic field generating an 
Aharonov-Bohm (AB) phase (see Fig.~\ref{pbs}). Due to the Rashba effect, spin up and spin down
carriers will pick up different phases along the upper arm and their interference pattern at the 
second beam splitter will be different. Choosing an appropriate phase difference we can ensure
 that at the second beam splitter a spin up (down) electron will always exit in the 0' (1') 
mode with unit probability \cite{MZ_buttiker}.

In order to calculate the transition amplitudes for each mode, we need to know the unitary 
transformations performed by each component. These are analyzed in the following. 
A beam-splitter acts only on the orbital degrees of freedom and is described by a symmetric
 $U(2)$ matrix:
\be
{\rm BS:}\ \ \ \ \ \ \ket{\sigma; k} \ra\ \cos\theta\, \ket{\sigma; k} + i\sin\theta\, \ket{\sigma; 1-k}
\label{bs}
\ee
where $t\equiv \cos\theta$ ($r\equiv i\sin\theta$) is the transmission (reflection)
 amplitude (the reflected component acquires a $\pi/2$ phase relative to the transmitted one). 
We denote by $\theta_{1,2}$ the parameters describing the two beam splitters.

Suppose we have an electron moving with velocity $\bf v$ in a region with a {\em static} 
electric field $\bf E$. Then the electron sees an effective magnetic field $\bf B \sim v
 \times E$ which couples to its spin. The spin-orbit (Rashba) Hamiltonian due to this coupling
 is $H_R\sim (\bf \hat{p} \times E)\cdot \overrightarrow\sigma$. In the mesoscopic context we 
are considering, the electron is
 confined to move in the $xz$ plane of a two dimensional electron gas (2DEG). We consider an 
electric field perpendicular to this plane, ${\bf E}= (0, E_y,0)$, which can be controlled by
 top/bottom gates \cite{grundler, spin_orbit}. Since in the Rashba region the electron is moving in the $x$ direction (see Fig.~\ref{pbs}), the Hamiltonian reduces to $H_R= \alpha \hat{p}_x \sigma_z/\hbar$ ($\alpha$ is the spin-orbit coupling which includes the effect of the applied field $\bf E$). 
The corresponding unitary transformation on the electronic wave-function is a rotation in the spin space given
 by $U_R= e^{i\phi_R\sigma_z}$, with $\phi_R= \alpha m^* L/\hbar^2$; $m^*$ is the effective 
electron mass and $L$ the length of the Rashba region. Note that the phase shift does not
 depend on the momentum of the incoming spin and hence it is non dispersive (this is correct 
if the interband coupling is negligible, which is true if the channel width $w \ll \hbar^2/\alpha 
m^*$ \cite{datta_das, egues}). Since it affects only the 1-mode (there is no Rashba coupling 
on the 0-mode), the transformation can be written as:
\be
{\rm Rashba:}\ \ \ \ \ \ket{\up; k} \ra\ e^{ik\phi_R}\ket{\up; k} \ \ , \ \ \ket{\down; k} \ra\ 
e^{-ik\phi_R}\ket{\down; k}
\label{rashba}
\ee
$k=0,1$. It is important to note that the quantization axis $Oz$ is defined as the in-plane direction
 perpendicular to the electron's wave-vector in the Rashba region. An in-plane rotation of 
this axis can be viewed as equivalent to a rotation of the $Oz$ axis of a Stern-Gerlach 
apparatus.

The magnetic flux $\Phi$ threading the interferometer generates an Aharonov-Bohm phase, which induces a phase difference in the electronic wave-functions between the two arms. Without loss of generality, we choose this phase to be on the upper arm (i.e., mode 1):
\be
{\rm AB:}\ \ \ \ \ \ \ket{\sigma; 0} \ra\ \ket{\sigma; 0} \ \ , \ \ \ket{\sigma; 1} \ra\ 
e^{i\phi_{AB}}\ket{\sigma; 1}
\label{ab}
\ee
where the AB phase is $\phi_{AB}=\Phi/\Phi_0$ ($\Phi_0= \hbar c/e$). It is important that 
the AB flux is confined to the center of the interferometer such that ${\bf B}= 0$ on the
 electron's path (otherwise the magnetic field will induce a spin precession).

In the described setup it is essential that carriers are charged particles with spin; the AB (Rashba) Hamiltonian
 couples to the charge (spin) degrees of freedom. For neutral particles with spin, the interferometer still works provided that we replace
 the 
magnetic flux $\Phi$ producing the AB phase with a potential well on one arm; this will induce 
a phase difference between the two paths, but in this case the phase is dispersive (likewise,
 the same phase difference can be induced if the two arms have different lengths).

From eqs.~(\ref{bs})-(\ref{ab}), we can now calculate the unitary transformation
 (the scattering matrix) performed by the whole device on the electronic wave-function. 
We assume that spins are injected only in one input (say 0) and the other input (e.g.~1)
 is kept free. Although only one input is used, both of them are required for preserving 
the unitarity of the device (seen as a {\sf CNOT} gate between the spin and the charge 
\cite{oneinput}). Since none of the interactions (\ref{bs})--(\ref{ab}) flips the spin, the 
unitary transformation performed by the device is (for simplicity we omit the ' on the output
 modes): $\ket{\up; 0} \ra\ t^\up_0\, \ket{\up; 0}+ t^\up_1\, \ket{\up; 1}$ and $\ket{\down; 0} \ra\ t^\down_0\, \ket{\down; 0}+ t^\down_1\, \ket{\down; 1}$, with $t^\sigma_k$ the corresponding transition amplitudes:
\begin{eqnarray}
\nonumber
\label{t0}
t^{\up,\down}_0= e^{i(\phi_{AB}\pm \phi_R)/2} \left[ \cos\frac{\phi_{AB}\pm \phi_R}{2} \cos(\theta_1+\theta_2)- \right. \\
\left. -i\sin\frac{\phi_{AB}\pm \phi_R}{2}\cos(\theta_1- \theta_2) \right] \\
\nonumber
\label{t1}
t^{\up,\down}_1= e^{i(\phi_{AB}\pm \phi_R)/2} \left[ i\cos\frac{\phi_{AB}\pm \phi_R}{2} \sin(\theta_1+\theta_2)- \right. \\
\left. -\sin\frac{\phi_{AB}\pm \phi_R}{2} \sin(\theta_1- \theta_2) \right]
\end{eqnarray}
Unitarity implies $\sum_k |t^\up_k|^2= \sum_k |t^\down_k|^2= 1$ (current conservation). Choosing $\theta_1=\theta_2=\pi/4$ (corresponding to 50/50 beam splitters) and $\phi_{AB}= \phi_R= \pi/2$, we achieve the desired transformation for a polarizing beam splitter: $\ket{\up; 0} \ra\ \ket{\up; 0}$, $\ket{\down; 0} \ra\ i\, \ket{\down; 1}$. Thus, a spin up is always transmitted in the same mode, whereas a spin down is always reflected in the opposite mode (up to a spurious phase which can be ignored or easily corrected) \cite{input1}.

In an experimental implementation of the proposed device, both phase $l_\phi^c$ and spin 
coherence length $l_\phi^s$ should be larger than the device size. For electrons at low 
temperatures, values of $l_\phi^c\sim 20\mu$m (at 15 mK) \cite{phase_coherence} 
and $l_\phi^s\sim 100\mu$m (1.6 K) \cite{spin_coherence} are reported in GaAs heterostructures. 
For carbon nanotubes (CNTs), $l_\phi^c\sim 1\mu$m at {\em room temperature} were observed 
\cite{ballistic_CNT}. Spin coherent transport ($l_\phi^s> 130\,$nm at 4.2 K) in CNTs has also been
 reported \cite{spin_CNT}. We can estimate the length $L$ of the Rashba region necessary for 
a rotation angle $\phi_R =\pi/2$ as $L= 58\,$nm in InAs ($\alpha= 4\times 10^{-11}$eVm 
\cite{grundler}) or $L= 250\,$nm in InGaAs/InAlAs ($\alpha= 0.93\times 10^{-11}$eVm 
\cite{spin_orbit}). A possible experimental implementation of the proposed device
 could exploit the 2DEG formed at the interface between a InAlAs and a InGaAs layer \cite{nitta}.

At this point we would like to make some remarks. Due to its topological nature, the AB
 phase has an important property, namely it is non-dispersive \cite{nondispersiveAB}. 
Hence, the AB phase acquired by an electron does not depend on its energy, or on the fact 
that it is monochromatic or not. The same property holds for the Rashba phase if interband coupling is negligible, as discussed above. Moreover, if the interferometer is balanced 
(the two arms have equal length) there will be no extra phase difference between the two 
paths. The beam splitters can also be considered non-dispersive, as pointed out 
in \cite{buttiker} (in that case the 50/50 beam-splitter is simply an intersection of two
 ballistic wires). Therefore the whole device will be non-dispersive, i.e.~the interference
 pattern will not depend on the energy of the incoming electrons or on the fact that they 
can be described by a wave-packet or a plane wave \cite{difference}. 

In practice, however, it is likely that the two arms will have different lengths
 ($l_0\ne l_1$) and this will induce an extra phase difference $\sim (l_0- l_1)/\lambda$, 
which is clearly dispersive ($\lambda$ is the electron wavelength). Experimentally it is then important to carefully calibrate the interferometer such that $l_0= l_1$. For a GaAs-based heterostructure the device could be designed by negatively biased metallic gates which would deplete the 2DEG underneath \cite{deplete}. In many mesoscopic experiments, the channel width can be varied by tuning the depletion gate voltage. Similarly, we can laterally shift the channel (keeping its width constant) by varying the potential difference between two gates. Thus, using an appropriate design it is possible to vary
 (within some limits) the length of one arm, such that in the end the interferometer is balanced. In order to minimize the errors electrons with $\lambda \gg |l_0-l_1|$ should be used.

In order to discuss the general properties of the device, it is convenient to make the following change of variables: $\epsilon_i\equiv \theta_i-\pi/4$, $\delta_{AB}\equiv \phi_{AB}-\pi/2$ and $\delta_R= \phi_R-\pi/2$, such that $\epsilon_i=\delta_{AB}=\delta_R=0$ corresponds to the ideal case. If the interferometer is not balanced ($l_0\ne l_1$), the extra phase arising can be always included in the AB phase $\delta_{AB}$ (however, in this case the total phase will be dispersive).

We define the spin polarization for the $k$ output as $P_k= (g^\up_k- g^\down_k)/(g^\up_k+ g^\down_k)$, where $g^\sigma_k=e^2\, |t^\sigma_k|^2/h$ is the conductance for spin $\sigma$ in mode $k$. From eqs.~(\ref{t0})--(\ref{t1}) we obtain:
\begin{eqnarray}
\label{p0p1}
P_0= \frac{A}{B+1}\ \ \ , \ \ \ P_1= \frac{A}{B-1}\label{pol}\\
A= \cos\delta_{AB} \cos\delta_R \cos 2\epsilon_1 \cos 2\epsilon_2\\
B= \sin 2\epsilon_1 \sin 2\epsilon_2- \sin\delta_{AB} \sin\delta_R \cos 2\epsilon_1 \cos 2\epsilon_2\label{B}
\end{eqnarray}
The ideal interferometer ($\epsilon_i=\delta_{AB,R}=0$) has $P_k=(-1)^k$, i.e., there is a totally spin up (down) polarized current in output 0 (1).

We can also define the efficiency of the spin $\sigma$ polarized current in mode $k$ as $\eta^\sigma_k\equiv g^\sigma_k/(g^\sigma_0+ g^\sigma_1)= |t^\sigma_k|^2$. Due to current conservation, only two of $\eta^\sigma_k$ are independent, say $\eta^\up_0$ and $\eta^\down_1$; then $\eta^\up_1= 1-\eta^\up_0$ and $\eta^\down_0= 1-\eta^\down_1$. From eqs.~(\ref{t0})--(\ref{t1}) we obtain:
\begin{eqnarray}
\eta^\up_0= (A+B+1)/2\\
\eta^\down_1= (A-B+1)/2
\end{eqnarray}
Note that in general the two outputs are not symmetric due to the asymmetry introduced by the Rashba interaction. Thus, it is possible to have situations in which the spin current is 100\% polarized in one output, but not in the other. This can happen, for example, if there is no spin down current in one output and the spin up current splits into both outputs. From (\ref{pol})--(\ref{B}) we can derive the conditions under which at least one output is completely polarized:\\
(a) $P_k=(-1)^k$ iff $\{ \epsilon_1= (-1)^{k+1}\epsilon_2,\ \ \delta_{AB}= (-1)^k\delta_R \}$ or $\{ \epsilon_1= (-1)^k\epsilon_2\pm \pi/2,\ \ \delta_{AB}= \pi+ (-1)^k\delta_R \}$;\\
(b) $P_k= (-1)^{k+1}$ iff $\{ \epsilon_1= (-1)^{k+1}\epsilon_2,\ \ \delta_{AB}= \pi-(-1)^k\delta_R \}$ or $\{ \epsilon_1= (-1)^k\epsilon_2\pm \pi/2,\ \ \delta_{AB}= (-1)^{k+1}\delta_R\}$,\\
with $k=0,1$.

It is important to note that in all four cases the efficiency of the completely polarized output is the same 
\be
\eta= \cos^2 \delta_R \cos^2 2\epsilon_2
\label{eta}
\ee
while the polarization of the other output (which in general is not completely polarized) is given by $P_{1-k}= P_k\,\eta/(\eta-2)$ ($P_k=\pm 1$). For the ideal interferometer, the efficiency attains its maximum $\eta=1$ and we recover $P_{1-k}= -P_k$: the two outputs have opposite polarizations (and 100\% efficiency). This shows that there is a whole class of parameters for which a complete spin polarized current can be obtained in (at least) one of the outputs, although with the smaller efficiency given by (\ref{eta}) compared to the ideal device (which has a unit efficiency in {\em both} outputs).

We now study how robust is the device against perturbations. For small deviations from the ideal values, we can expand the polarizations (\ref{p0p1}) up to second order to obtain 
\begin{eqnarray}
P_0= 1- 2(\epsilon_1+ \epsilon_2)^2- (\delta_{AB}- \delta_R)^2/2+ {\cal O}(x^3)\\
P_1= -1+ 2(\epsilon_1- \epsilon_2)^2+ (\delta_{AB}+ \delta_R)^2/2+ {\cal O}(x^3)
\end{eqnarray}
This shows that the device is quite robust against small fluctuations, since the leading
 correction to the ideal result is quadratic. This is to be expected, since $\epsilon_i=\delta_{AB}=\delta_R=0$ is a stationary point at which $P_0\, (P_1)$ reaches its maximum (minimum).

\begin{figure}
\putfig{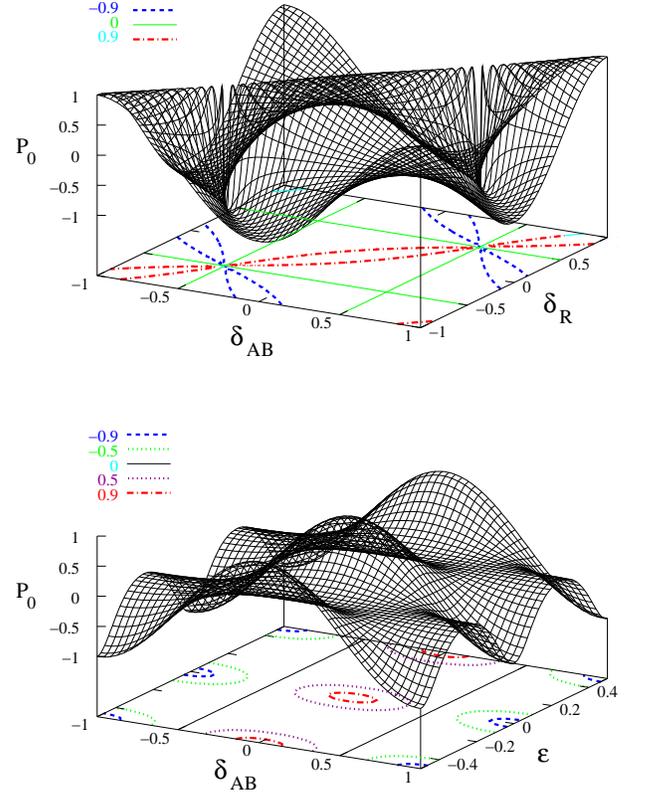}{8}
\caption{Upper panel: $P_0$ as a function of $\delta_{AB}$ and $\delta_R$ (both in $\pi$ units) and $\epsilon_{1,2}=0$. The contour lines correspond to $P_0=0.9$ (dashed line), $P_0=-0.9$ (dashed-dot line) and  $P_0=0$ (solid line).
Lower panel: $P_0$ versus $\delta_{AB}$ and $\epsilon\equiv\epsilon_1=\epsilon_2$ (in $\pi$ units) with $\delta_R=0$. The contour lines correspond to $P_0=0.9$ (dashed line), $P_0=-0.9$ (dashed-dot line), $P_0=\pm0.5$ (dotted line) and $P_0=0$ (solid line).}
\label{polar0}
\end{figure}

In the upper panel of Fig.~\ref{polar0} we plot $P_0$ as a function of $\delta_{AB}$ and $\delta_R$, with $\epsilon_{1,2}=0$. 
In the lower panel we show $P_0$ as a function of $\delta_{AB}$ and $\epsilon\equiv\epsilon_1=\epsilon_2$ for $\delta_R=0$. We obtain the same result if we interchange $\delta_{AB}$ with $\delta_R$. The contour lines for $P_0=\pm 0.9$ indicates that in both cases there are relevant regions in the parameter space in which the output polarization is greater than 90\%. Fig.~\ref{polar0} suggests that in order to obtain a spin polarization close to unity, a deviation from the ideal value of one variable can be compensated by tuning in a clever way one of the other parameters, such as the applied magnetic or electric field.

{\em Applications.} The PBS described in this letter can be used in several setups. It can be
 used as a {\em preparation} device to produce spin polarized electrons with high 
(theoretical 100\%) efficiency. It can also be used as a {\em measuring} device, since it 
is the mesoscopic equivalent of a Stern-Gerlach apparatus. Direct measuring of spin in 
a mesoscopic context is difficult, one of the problems being that spin filtering techniques 
are not efficient. Moreover, in some quantum computation schemes \cite{barnes}, spins have
 to be measured individually, a task difficult to achieve (filtering cannot be used, since 
it will imply absorption of some of the spins). We stress that a PBS converts spin degrees of freedom into orbital (charge) degrees of freedom and therefore {\em single} spin detection becomes feasible in this scheme by using single electron transistors (SETs) coupled to the output modes.

In a spintronic context, where detection of individual spins is not required, a ratio of the two spin polarized output currents will give information about the polarization of the input current. Suppose that the input state is in a spin superposition $\cos\theta \ket{\up; 0} + \sin\theta \ket{\down; 0}$. Then, the ratio of the (spin polarized) output currents will be $I_{1'}/I_{0'}= \tan^2 \theta$.

In conclusion, we have proposed an interferometric device capable to separate an incoming unpolarized current into two totally polarized currents. Since no ferromagnetic contacts are used, the device architecture is simplified and an all semiconductor implementation is thus possible.

\noindent {\bf Acknowledgments.} We are grateful to Ehoud Pazy and Fabio Taddei for useful comments and enlightening discussions.


\begin{thebibliography}{}

\bibitem{spintronics} S.A.~Wolf {\it et al.}, Science {\bf 294}, 1488 (2001).

\bibitem{datta_das} S.~Datta, and B.~Das, \apl {\bf 56}, 665 (1990).

\bibitem{Schmidt:2000} G.~Schmidt, D.~Ferrand, L.W.~Molenkamp, A.T.~Filip and B.J.~van Wees \prb {\bf 62}, R4790 (2000).

\bibitem{Schmidt:2002} G.~Schmidt, C.~Gould, P.~Grabs, A.M.~Lunde, G.~Richter, A.~Slobodskyy, and L.W.~Molenkamp, cond-mat/0206347.

\bibitem{Flatte:20022002} Z.G.~Yu and M.E.~Flatt\'e, cond-mat/0201425; Z.G.~Yu and M.E.~Flatt\'e, cond-mat/0206321.

\bibitem{Rashba} E.I.~Rashba, cond-mat/0206129, Euro.~Phys.~J. B (to be published).

\bibitem{spin_injection} P.R.~Hammar {\it et al.}, \prl {\bf 83}, 203 (1999).

\bibitem{gardelis} S.~Gardelis {\it et al.}, \prb {\bf 60}, 7764 (1999); F.G.~Monzon {\it et al.}, \prl {\bf 84}, 5022 (2000); B.J.~van Wees, {\em ibid.} {\bf 84}, 5023; P.R.~Hammar {\it et al.}, {\em ibid.} {\bf 84}, 5024.

\bibitem{LaBella-Byers} V.P.~LaBella {\it et al.}, Science {\bf 292}, 1518 (2001); W.F.~Egelhoff Jr. {\it et al.}, Science {\bf 296}, 1195a (2002); V.P.~LaBella {\it et al.}, Science {\bf 296}, 1195a (2002). 

\bibitem{spin_injection2} R.~Fiederling {\it et al.}, Nature {\bf 402}, 787 (1999).

\bibitem{ohno} Y.~Ohno {\it et al.}, Nature {\bf 402}, 790 (1999).

\bibitem{jonker} B.T.~Jonker {\it et al.}, \prb {\bf 62}, 8180 (2000).

\bibitem{t_spin} A.A.~Kiselev, and K.W.~Kim, cond-mat/0203261.

\bibitem{t_spin1} A.A.~Kiselev, and K.W.~Kim, \apl {\bf 78}, 775 (2001).

\bibitem{spin_filter} M.~Governale, D.~Boese, U.~Z\"ulicke, and C.~Schroll, \prb {\bf 65}, 140403 (2002); cond-mat/0108373.

\bibitem{spin_switch} D.~Frustaglia, M.~Hentschel, and K.~Richter, \prl {\bf 87}, 256602 (2001).

\bibitem{QDspin_filter} P.~Recher, E.V.~Sukhorukov, and D.~Loss, \prl {\bf 85}, 1962 (2000).

\bibitem{nitta} J.~Nitta, F.E.~Meijer, and H.~Takayanagi, \apl {\bf 75}, 695 (1999).

\bibitem{oneinput} We point out that even in the case in which a single-input design is considered (e.g., by replacing the first beam splitter with a Y-junction), the corresponding device can still be used as a spin polarizer and all the results derived here for the output spin-polarization apply as well. In this case, however, the device is no longer equivalent to a PBS or a {\sf CNOT} gate and for this reason we prefer the four terminal design.

\bibitem{MZ_buttiker} A Mach-Zehnder interferometer was also considered by Seelig and B\"uttiker \cite{buttiker} in the context of charge decoherence in a AB ring.

\bibitem{buttiker} G.~Seelig, and M.~B\"uttiker \prb {\bf 64}, 245313 (2001).

\bibitem{grundler} D.~Grundler, \prl {\bf 84}, 6074 (2000).

\bibitem{spin_orbit} J.~Nitta {\it et al.}, \prl {\bf 78}, 1335 (1997).

\bibitem{egues} J.C.~Egues, G.~Burkard, and D.~Loss, cond-mat/0207392.

\bibitem{input1} If we inject spins in input 1 instead of 0, the transformations are: $\ket{\up; 1} \ra\ -\ket{\up; 1}$, $\ket{\down; 1} \ra\ i\, \ket{\down; 0}$.

\bibitem{phase_coherence} G.~Cernicchiaro {\it et al.}, \prl {\bf 79}, 273 (1997).

\bibitem{spin_coherence} J.M.~Kikkawa, and D.D.~Awshalom, Nature {\bf 397}, 139 (1999).

\bibitem{ballistic_CNT} A.~Bachtold {\it et al.}, \prl {\bf 84}, 6082 (2000); cond-mat/0002209.

\bibitem{spin_CNT} K.~Tsukagoshi, B.W.~Alphenaar, and H.~Ago, Nature {\bf 401}, 572 (1999).

\bibitem{nondispersiveAB} G.~Badurek {\it et al.}, \prl {\bf 71}, 307 (1993).

\bibitem{difference} This is one of the main differences between our proposal and the one described in \cite{t_spin}. Since in that case the interferometer is not balanced, the interference pattern, and hence the transmission coefficients, are both energy dependent.

\bibitem{deplete} E.~Bucks {\it et al.}, Nature {\bf 391}, 871 (1998).

\bibitem{barnes} C.H.W.~Barnes, J.M.~Shilton, and A.M.~Robinson, \prb {\bf 62}, 8410 (2000); cond-mat/0006037.

\end{thebibliography}
\end{document}